\documentclass{osa-article}

\journal{osac}

\articletype{Research Article}

\newcommand{\rb}{\right)}
\newcommand{\lb}{\left(}

\usepackage{lineno}

\begin{document}

\title{The role of delay-times in delay-based Photonic Reservoir Computing}

\author{Tobias H\"{u}lser,\authormark{1} Felix K\"{o}ster,\authormark{1} Lina Jaurigue,\authormark{1} and Kathy L\"{u}dge\authormark{2}}

\address{\authormark{1} Institute of Theoretical Physics, Technische Universi\"{a}t Berlin, Hardenbergstr. 36, 10623 Berlin, Germany\\
\authormark{2}Institute of Physics, Technische Uinversit\"{a}t Ilmenau, Weimarer Str. 25, 98693 Ilmenau, Germany}

\email{\authormark{*}kathy.luedge@tu-ilmenau.de} 
\begin{abstract}
Delay-based reservoir computing has gained a lot of attention due to the relative simplicity with which this concept can be implemented in hardware. However,there is still an misconception about the relationship between the delay-time and the input clock-cycle which has noticeable consequences for the performance. We review the existing literature on this subject and introduce the concept of delay-based reservoir computing in a manner that demonstrates that there is no predefined relationship between these two times-scales. Further, we discuss ways to improve the computing performance of a reservoir formed by delay-coupled oscillators and show the crucial impact of delay-time tuning in those multi-delay systems. 
\end{abstract}

\section{Introduction}

Reservoir Computing is a human brain-inspired machine learning
scheme that is versatile, has fast training times, and utilises the intrinsic information-processing capacities of dynamical systems \cite{JAE01,MAA02}. The simple training scheme that is employed in reservoir computing, which involves only training the output layer, avoids difficulties that typically arise in the training of recurrent neural networks, such as the vanishing gradient in time \cite{HOC98}. This makes reservoir computing particularly suited towards hardware implementation \cite{DOC09,ANT16,VAN14,FER03}. In the context of optimising computation speeds and power consumption, optical implementations are of particular interest and were made feasible by the introduction of so-called delay-based reservoir computing in \cite{APP11}. In \cite{APP11} the authors drew the connection between time-delayed systems and networks to show that a single non-linear node with time-delayed feedback could serve as a reservoir in which the reservoir responses are multiplexed in time rather than in space. Various experimental implementations of this delay-based scheme have been realised, including opto-electronic \cite{APP11,LAR12,PAQ12,CHE19c}, optical \cite{BRU13a,DEJ14,VIN15,HOU18} and electromechanical \cite{DIO18} systems, and delay-based reservoirs have demonstrated good performance in applications including time-series-predictions \cite{BUE17,KUR18}, fast word recognition \cite{LAR17} and signal conditioning \cite{ARG20}.

Despite the extensive research that has been carried out on delay-based reservoir computers since their inception in 2011 \cite{APP11}, there has been very little focus on the influence of the delay-times on the reservoir computing performance. In most studies one of two fixed values are chosen for the delay-time; either resonance with the input clock-cycle \cite{APP11,ORT17a,TAK18,SUG20} or slightly larger than the input clock-cycle, commonly referred to as the desynchronised regime \cite{PAQ12,DEJ14}. Furthermore, it is often stated that it is necessary for the delay-based reservoir computing concept for the delay-time to be chosen as one of these values. This is a misconception that stems from viewing delay-based reservoirs as networks. In this article, we will introduce the concept of delay-based reservoir computing in a manner that demonstrates that there is no predefined relationship between the delay-times of the reservoir and the input clock-cycle. We will review delay-based reservoir computing in the context of the influence of the delay-times, high-lighting the effect of resonances between the reservoir delay-times and the input clock-cycle on the computing performance.

We start  with reviewing the effect of tuning the delay in single-delay reservoir computing systems and subsequently show, how adding more delays allows for an even better tunability of the reservoir properties and thus leads to performance improvement. Besides some general rules that have to be obeyed in order to improve the results, e.g. avoiding resonances with the input clock-cycle, additional delays also add additional timescales that can be relevant for specific tasks. In systems with delay, resonances between delays and other characteristic time-scales of a given system generally play an important role and delay-related memory and resonance effects are largely universal. 

This work is structured as follows. First, we will give a brief introduction to the general reservoir computing concept in Sec.~\ref{Sec:RC}. In Sec.~\ref{Sec:DRCconcept} we will introduce delay-based reservoir computing by first introducing the concept of time-multiplexed reservoir computing and then explaining why time-delay systems serve as suitable time-multiplexed reservoirs. Following this, we will review the existing literature on aspects of this topic in Sec.~\ref{Sec:3.2}-\ref{Sec:3.4}, Sec.~\ref{sec:4} and \ref{sec:5}. Then, in Sec.~\ref{Sec:new}, we present our results for the two delay-coupled oscillators, where we show how the memory capacity can be tuned, and how choosing the correct delays can lead to a substantial performance improvement for the NARMA10 task.

\section{Reservoir computing concept}\label{Sec:RC}

\begin{figure}
    \centering
    \includegraphics[width=0.5\textwidth]{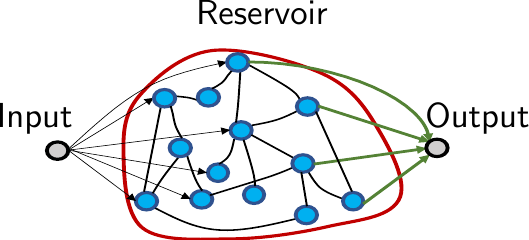}
   \caption{Sketch of a reservoir computer based on a recurrent neural network (echo state machine). The green lines represent the connections that need to be trained while the rest of the network remains unchanged.}
   \label{fig:RC}
\end{figure}

There are many articles and reviews that give good and thorough introductions to the reservoir computing concept. Here we will give a brief introduction and refer the reader to \cite{JAE01,LUK09,SAN17a,BRU18,TAN19c} for more in-depth discussions of the concepts and mathematical foundations.

A reservoir computer consists of three parts; the input layer, the reservoir, and the output layer. Figure \ref{fig:RC} shows a sketch of this layout for the case of one-dimensional input and output layers (the concept can easily be extended to high-dimensional inputs and outputs \cite{BRU18a}). In this sketch, we have depicted the reservoir as a recurrent neural network, as this was the original system for which the reservoir computing concept was conceived and allows for a direct comparison between reservoir computing and more traditional machine learning algorithms. However, we would like to emphasize that the reservoir does not need to be a network for reservoir computing to work. It is the training method that sets reservoir computing apart from other machine learning methods. Specifically, only the weights connecting the reservoir with the output layer are trained (connections highlighted in green in Fig.~\ref{fig:RC}). The input weights (connections between the input layer and the reservoir) and any internal properties of the reservoir are kept fixed during training. Typically, the input weights are chosen randomly, as are the internal weights when the reservoir is a recurrent neural network.

In the training phase, the reservoir is fed $K_\textrm{tr}$ successive inputs and the response of the reservoir is sampled $N$ times for each input. These responses (node states in the case of a neural network reservoir) are written into an $K_\textrm{tr}\times \left(N+1\right)$-dimensional state matrix ${\bf \underline{\underline{S}}}$, the last column of which is filled with a bias term of one. The training step is then to find the weights ${\bf \underline{W}^\textrm{out}} \in \mathbb{R}^{N+1}$ that minimise the difference between the output ${\bf \underline{o}}={\bf \underline{\underline{S}}}^{~}{\bf \underline{W}^\textrm{out}} \in \mathbb{R}^{K_\textrm{tr}}$ and the target output ${\bf \hat{\underline{o}}}\in \mathbb{R}^{K_\textrm{tr}}$, i.e. the ${\bf \underline{W}^\textrm{out}}$ that solves
\begin{align}\label{gl:minimierungsproblem_lineare_regression}
  \min_{{\bf \underline{W}^\textrm{out}}}\left(||{\bf \underline{\underline{S}}}^{~} {\bf \underline{W}^\textrm{out}} - {\bf \hat{\underline{o}}}||^2_2 +\lambda||{\bf \underline{W}^\textrm{out}}||^2_2\right),
\end{align}
where $\lambda$ is the Tikhonov regularisation parameter and $||\cdot ||_2$ is the Euclidean norm. The solution to this problem can be found using the Moore-Penrose pseudoinverse and is given by
\begin{align}\label{Eq:wout}
    {\bf \underline{W}^\textrm{out}}=({\bf \underline{\underline{S}}}^T{\bf \underline{\underline{S}}}+\lambda {\bf \underline{\underline{I}}})^{-1} {\bf \underline{\underline{S}}}^T {\bf \hat{\underline{o}}}.
\end{align}
For the case of a multi-dimensional output, the target sequence and the output weight vectors become matrices and Eq.~(\ref{Eq:wout}) is adjusted accordingly \cite{BRU18a}.

\subsection{Performance measure}

To quantify the quality of the prediction, ${\bf \underline{o}}={\bf \underline{\underline{S}}}^{~}{\bf \underline{W}^\textrm{out}}$, there are various error measures (see for example \cite{BOT19,FLA19}). One commonly used measure is the normalised root-mean-square error (NRMSE), defined as
\begin{equation}
 \textrm{NRMSE}=\sqrt{\frac{\sum_{k'=1}^{K_{o}}\lb \hat{o}_{k'}-o_{k'}\rb^2}{K_{o} \textrm{var}\lb {\bf \underline{\hat{o}}}\rb}},
\end{equation}
where $\hat{o}_{k'}$ are the target values, $o_{k'}$ are the outputs produced by the reservoir computer, $K_{o}$ is the length of the vector ${\bf \underline{\hat{o}}}$ and $\textrm{var}\lb {\bf \underline{\hat{o}}} \rb$ is the variance of the target sequence.

\section{Delay-based reservoir computing}
\subsection{Concept}\label{Sec:DRCconcept}

\begin{figure}
    \centering
    \includegraphics[width=\textwidth]{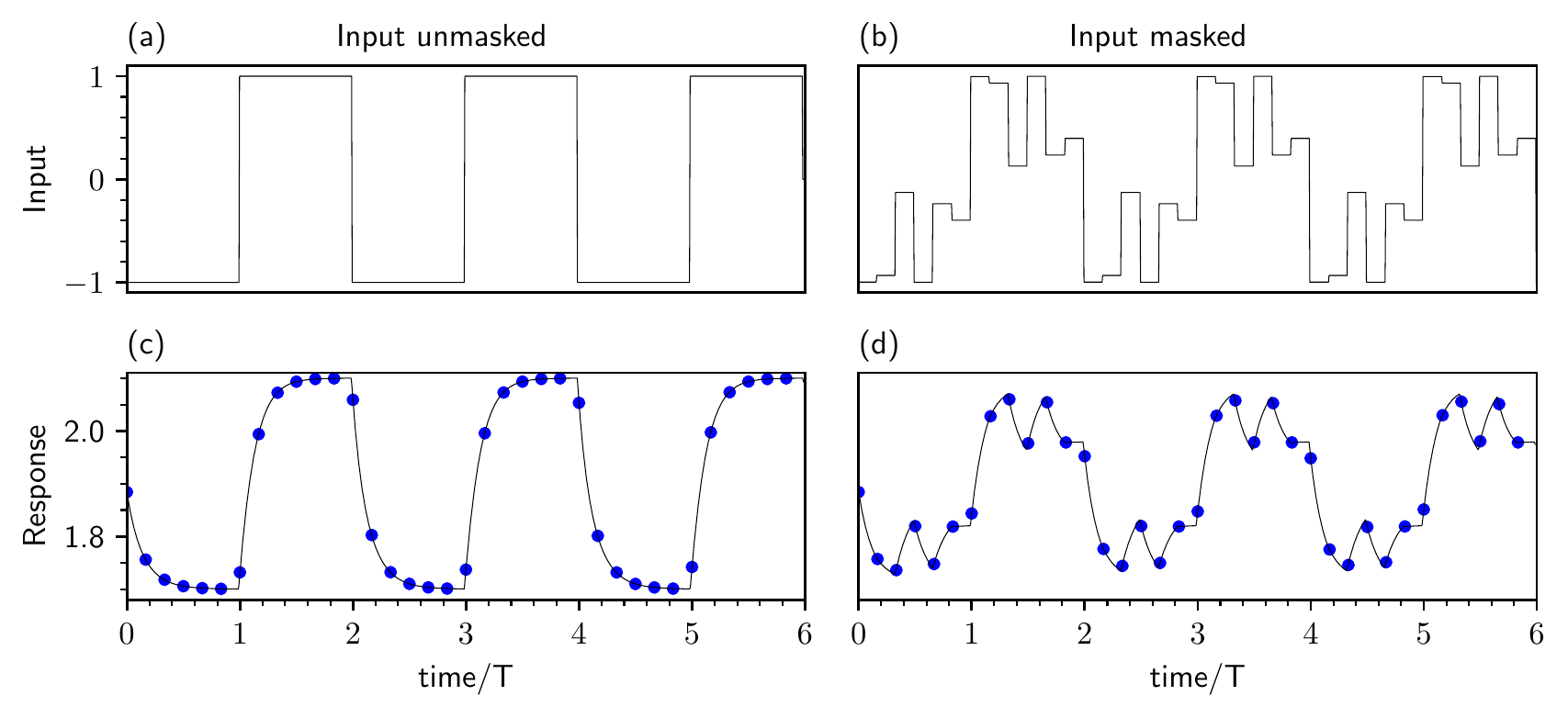}
   \caption{Visualisation of time-multiplexing (sequential sampling indicated by blue dots) and masking. Left and right panel show unmasked and masked input (top row) while the respective system response is shown below. Time is normalized to the input cycle $T$. }
   \label{fig:mask}
\end{figure}

We will introduce the concept of delay-based reservoir computing in a different manner to the usual approach. We do this to clarify a common misconception about the relationship between the delay-time $\tau$ of the reservoir and the input clock-cycle $T$, the misconception being that these quantities must be equal, or more generally, that there is any predefined relationship between these quantities. In order to do this, we will first introduce the concept of time-multiplexed reservoir computing independently of delay-based systems and will subsequently show why time-delayed systems are suitable reservoirs for time-multiplexed reservoir computing.

\subsubsection{Time-multiplexed reservoir computing}

One of the main concepts underlying delay-based reservoir computing is that rather than multiplexing in space, as is done in conventional reservoir computing, one multiplexes in time. This means that the response of the reservoir is sampled multiple times as a function of time rather than as a function of space. In such a scheme, each piece of input data must be fed into the reservoir for a certain time interval, which we will refer to as the clock-cycle $T$. During that interval, the response of the system must be sampled a number of times. A sketch of an example input sequence and the corresponding reservoir responses is shown in Fig.~\ref{fig:mask}a. The number of times that the system is sampled (blue circles in Fig.~\ref{fig:mask}a) gives the output dimension, commonly referred to as the virtual nodes $N_v$ in delay-based reservoir computing. It is analogous to the output dimension $N$ in the general reservoir computing scheme described in Section~\ref{Sec:RC}. 

In principle, any dynamical system can be used as the reservoir for such a time-multiplexed scheme, however, the dynamics and characteristic time-scales of the system play an important role in the performance that can be achieved. It has been shown that systems generally perform well when the dynamics of the underlying autonomous system (i.e. the system not driven by the input data) are steady state dynamics but the system is close to a bifurcation leading to chaotic or other high-dimensional dynamics (\cite{NAK21, CAR20b}). As indicated in Fig.~\ref{fig:mask}a, if the reservoir relaxes back to its steady state within an input clock-cycle, multiple identical system responses are sampled. As these responses are not linearly independent, the effective output dimension of the reservoir is reduced. One way to mitigate this problem is to apply a mask to the input data. An example of a masked input is shown in Fig.~\ref{fig:mask}b. By disturbing the system within one input cycle $T$, the response can be diversified. If the length of the mask steps, typically labelled $\theta$, is chosen appropriately the system can be prevented from relaxing to its steady state. 

Typically, step functions with randomly selected step heights are used for the masking procedure, but, work has also been done on the performance of chaotic, multilevel, binary, or pattered masks \cite{KUR18, APP14,SOR13a,ARG21}. In the example shown in Fig.~\ref{fig:mask}b randomly selected step heights are used and the system is sampled at the end of each mask step. The choice of sampling position is arbitrary, however, it must be the same in each clock-cycle. It has also been shown that it can be beneficial to vary the lengths of the mask step and the sampling positions within one clock-cycle \cite{GOL21a,YUE19,TOU15}. Which type of mask results in the best performance is both system and task-dependent.

\subsubsection{Time-multiplexed RC with a delay system}

\begin{figure}
    \centering
    \includegraphics[]{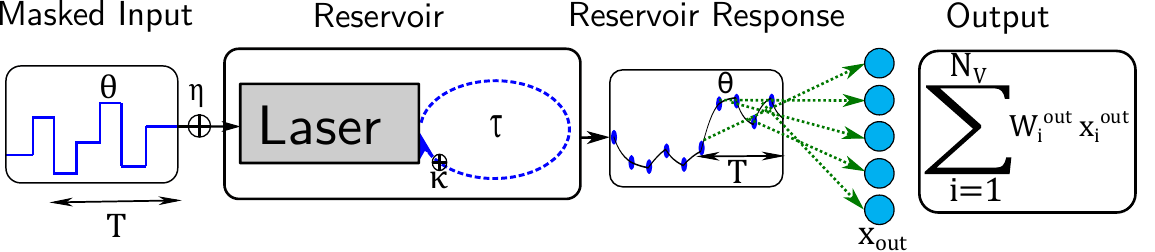}
   \caption{Sketch of a delay-based reservoir computer realized with a laser subjected to optical feedback, masked input, and time-multiplexed output.}
   \label{fig:delayRC}
\end{figure}

Having introduced the concept of time-multiplexed reservoir computing, we will now address why time-delayed systems are suitable as reservoirs in this scheme. For reservoir computing to yield good results, the responses of the reservoir to sufficiently different inputs must be linearly separable. For this to be possible the state-space of the reservoir must be sufficiently high-dimensional. This can be achieved by having a large network of coupled nodes or by using one nonlinear node with time-delayed feedback. For delay-based reservoir computing delay differential equations of the following form are often considered (see for example \cite{APP11, LAR17,BRU18a,ROE19,STE20}):
\begin{equation}\label{Eq:DDE}
    \frac{dx\left(t\right)}{dt}  =-x\left( t\right)+f \left[ x\left( t-\tau\right)+\eta J\left( t \right) \right],
\end{equation}
where $f \left[ \cdot \right]$ is a non-linear function, $J\left( t \right)$ is the input function and $\eta$ is the input scaling. In a strict mathematical sense, such a time-delayed continuous system is infinitely-dimensional \cite{HAL93} as the initial conditions of the system must be defined over the entire time interval $[-\tau,0]$. In a practical sense, time-delayed systems do not behave as infinite-dimensional systems, but they can exhibit complex dynamics such as quasi-periodicity and chaos. Close to bifurcations leading to these dynamics, time-delayed systems can also exhibit complex and high-dimensional transient dynamics and can therefore perform well as time-multiplexed reservoirs. Furthermore, the delay term introduces an additional, tunable, time scale that directly influences the memory capabilities of the reservoir (see Sec.~\ref{Sec:new}). Due to the relative simplicity with which time-delayed feedback can be implemented experimentally, especially in photonic systems, time-multiplexed reservoir computing using delayed-systems has been extensively researched since its introduction in \cite{APP11}.

In Fig.~\ref{fig:delayRC}, we show a sketch of a time-delayed reservoir. The system is fed a masked input sequence and the response of the system is sampled at predefined time intervals. Here we would like to emphasis that the time-multiplexing procedure can be viewed completely independently of the time-delayed system and there is no predefined relationship between the delay time $\tau$ and the clock-cycle $T$. However, resonances between $\tau$ and $T$ can have a significant impact on the performance that can be achieved, as will be discussed in Section~\ref{Sec:res}.

\subsection{Viewing delay-based reservoirs as networks}\label{Sec:3.2}

\begin{figure}
    \centering
    \includegraphics[width=0.8\textwidth]{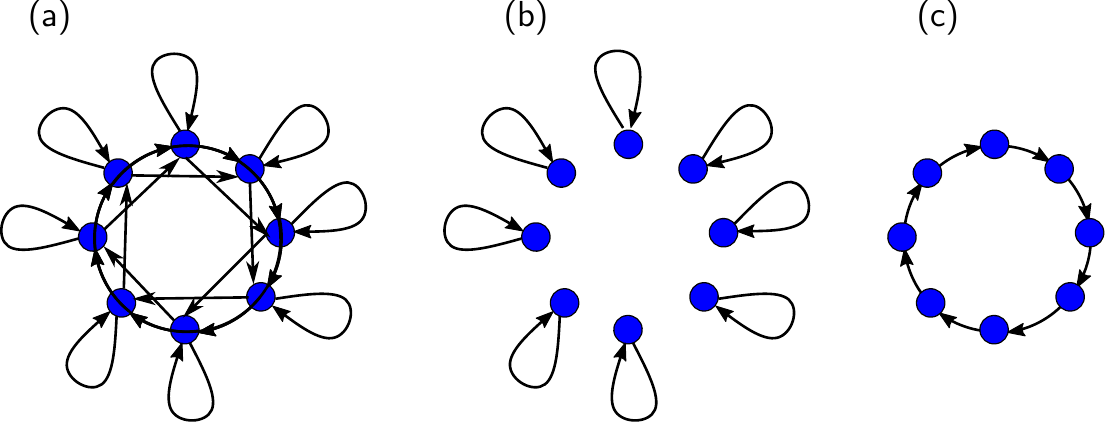}
   \caption{Network topologies as they arise in delay-based reservoir computing systems when the delay is restricted to $\tau=n \theta$ ($n\in \mathbb{N}$). In (a,b) the input cycle $T$ is chosen resonant to the delay,  $\tau=T$, while  $\tau=T+\theta$ in (c). The characteristic system time $T'$ is larger than $\theta$ in (a) while it is smaller than $\theta$ in (b,c). }
   \label{fig:networks}
\end{figure}

When the delay-based reservoir computing scheme was introduced in \cite{APP11} the authors introduced the concept of virtual nodes equidistantly distributed along the delay-line. They chose $T=\tau=\theta N_v$, where $N_v$ is the number of virtual nodes, and described how this system can be considered as a network. They considered the two cases $\theta < T'$ and $\theta >>T'$, where $T'$ is the characteristic timescale of the nonlinear node. In the first case, $\theta < T'$, the system is perturbed by each subsequent input step before it can relax to a steady state. This means that the transient state of the system for one input interval depends on its state in the previous intervals. This can be viewed as a unidirectional coupling between neighbouring virtual nodes. Figure~\ref{fig:networks}a shows a sketch of the network that such a system emulates. In addition to the coupling via the nonlinear node dynamics, choosing $\tau=\theta N_v$ means that in such a network view of the system, the virtual nodes are also coupled with their state in the previous clock-cycle, as indicated by the self-feedback loops in Fig~\ref{fig:networks}a.

In the second case, $\theta >>T'$, the nonlinear node relaxes to its steady state within one $\theta$ mask interval, this means that there is no dependence on the state of the system during the previous mask step and the system can be described by a set of independent nodes with self-feedback (see Fig.~\ref{fig:networks}b). In this scenario, for reservoirs of the form given by Eq.~(\ref{Eq:DDE}), the delay-differential equation describing the reservoir can be reduced to a set of recursive equations that describe the state of the virtual nodes. Let $x_k$ be $x\left( t\right)$ at the end of the $k^\textrm{th}$ $\theta$ interval of the piece-wise constant masked input function $J\left( t\right)$, i.e. $x_k=x\left( t\right)$ for $t=\left( k+1\right)\theta$ with $J\left( t\right)=J_k$ for $k\theta < t \leq \left( k+1\right)$. Then, if the dynamics of the underlying autonomous system are steady state dynamics, for $\theta >>T'$, Eq.~(\ref{Eq:DDE}) can be written as
\begin{equation}\label{Eq:map1}
    0 =-x_k+f\left[ x_{k-Nv}+\eta J_k\right].
\end{equation}
Rewriting $J_k$ in terms of the input data sequence $I_{k'}$ and the mask values $M_n$, and relabelling $x_k$ in terms of the virtual nodes number $n\in[0,N_v)$ and the input index $k'$, Eq.~(\ref{Eq:map1}) becomes
\begin{equation}\label{Eq:map2}
    x_{k'}^n=f\left[ x_{k'-1}^n+\eta M_n I_{k'} \right].
\end{equation}

If, instead of $\tau=\theta N_v$, the delay is chosen as $\tau=\theta \left(N_v+1\right)$, as was introduced in \cite{PAQ12}, then Eq.~(\ref{Eq:map2}) becomes 
\begin{equation}
    x_{k'}^n=f\left[ x_{k'-1}^{n-1}+\eta M_n I_{k'} \right] \quad \textrm{for} \quad n\in \left[1,N_v-1\right], ~~\textrm{and}\quad
    x_{k'}^0=f\left[ x_{k'-2}^{N_v-1}+\eta M_0 I_{k'} \right] .
\end{equation}
In this case, the network representation is equivalent to a unidirectional ring, as sketch in Fig.~\ref{fig:networks}c.

The equivalences between time-delayed systems and networks, and more generally the now well-known equivalence between time-delayed and spatially extended systems \cite{ARE92}, lead to the inception of delay-based reservoir computing. However, continuing to view delay-based reservoirs as networks puts unnecessary constraints on the relationship between the delay-time $\tau$ and the clock-cycle $T$. Mainly because the direct equivalence is only applicable when $\tau=n\theta$ for $n\in \mathbb{N}$ and $\theta=T/N_v$ \cite{STE20}. It is not necessary to view delay-based systems as networks for reservoir computing purposes and letting go of this idea could lead to improved performance. This is because the delay-time plays an important role in determining the memory of the reservoir and allowing this parameter to be tuned more freely could lead to better fulfilment of task specific memory requirements \cite{ROE19,KOE20a,KOE21}.

\subsection{Delay and clock-cycle resonances}\label{Sec:res}

It is generally known that resonances between various characteristic time-scales can have an important influence on the dynamics of a system. This is a phenomenon that is observed in a multitude of applications, including  neural oscillations \cite{PAN12,CAL17} or mode-locked lasers \cite{JAU16,JAU16a}. Sometimes these resonances are beneficial to the desired outcome and other times they have a detrimental effect. In the case of delay-based reservoir computing, resonances between the delay-time of the reservoir and the clock-cycle generally have a detrimental effect on the performance. This has been demonstrated in a number of publications \cite{ORT19,ROE19,STE20,KOE20a,KOE21,JAU21a}. An example of $\tau$-$T$ resonances is shown in Fig.~\ref{fig:res_plot}, wherein the memory capacity and the NARMA10 NRMSE (see the Supplemental Material for the definitions of these two commonly used benchmarking tasks) are plotted as a function of $\tau$ and $T$. In Fig.~\ref{fig:res_plot}a, dips in the memory capacity are found at the main resonance $\tau=T$ and at higher order resonances $p\tau=qT$ for $p,q\in \mathbb{N}$ and in Fig.~\ref{fig:res_plot}b increases in the NRMSE are found at some of these resonances for the NARMA10 task. By inspecting the evolution of the performance in Fig.~\ref{fig:res_plot}b, it also becomes clear that the optimal NARMA10 performance is not achieved at the commonly used relation of $\tau=T+\theta$.

\begin{figure}
    \includegraphics[width=1\textwidth]{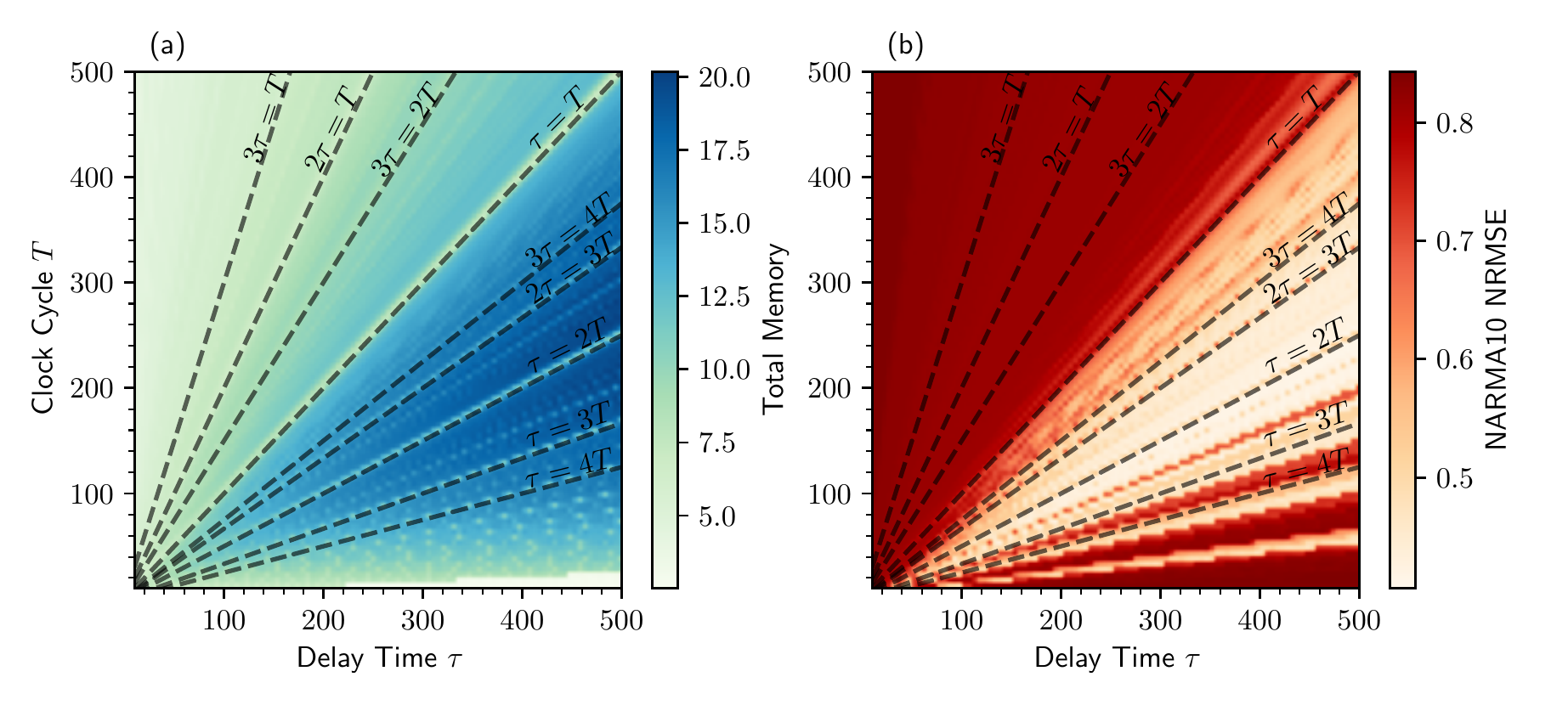}
   \caption{Total memory capacity (a) and performance of NARMA10 prediction (NRMSE) (b) plotted as a function of clock-cycle $T$ and delay-time $\tau$. Dashed lines indicate resonances of rational ratio between the two timescales, i.e. $n T = m \tau$, with $n,m \in \mathbb{N}$.}
   \label{fig:res_plot}
\end{figure}

Two explanations for this resonance phenomenon were given in \cite{STE20} and \cite{KOE21}. The first, taken from \cite{STE20}, translates the delay-based reservoir computer via an Euler-step method into an equivalent spatially extended network. This method works for a constrained set of delay differential equations.
Looking at the connections of this equivalent network system, one sees a reduction of node interconnection for resonant setups.
Thus, a resonant case yields a badly connected network, lowering the reservoir computation capabilities.

The second explanation, given in \cite{KOE21}, uses an eigenvalue analysis
and describes how the reservoir reacts to the inputs in the regime where the input is a small perturbation. They show that resonances between the delay-time $\tau$ and clock cycle $T$ yield a multitude of overlapping half-circle rotations in the many complex planes of the delay-based reservoir computers infinite-dimensional phase space.
As mentioned above, practically, the reservoir only has a finite number of accessible eigendirections, as all but a finite number of eigendirections decay infinitely fast.
Thus, the remaining finite eigendirections yield the usable phase space of the reservoir to project the input information into.
In the case of a long delay-time $\tau$ (long in comparison to the local timescale of the reservoir, which is the case in typical reservoir computing settings), the imaginary parts of all eigenvalues $\mu_{k}$ are approximately given by
\begin{align}
\mu_{k} \approx \frac{\pi}{\tau}(2k - \nu),
\end{align}
with $k \in \mathbb{N}$ being the index of the $k$-th eigenvalue and $\nu \in \{0,1\}$ being a constant phase shift.
Now, taking a look at the rotation of a trajectory in one clock cycle time $T$ in one of the $k$ complex planes, yields $T \mu_{k} \approx \frac{T \pi}{\tau}(2k - \nu)$.
This is the traversed angle of rotation during one input clock cycle $T$.
Setting the clock cycle $T$ to some rational part of the delay-time $T= n \tau$ (i.e. in resonance with $n \in \mathbb{Q}$), the traversed angle will always be a rational multiple of $\pi$, with $T \mu_{k} \approx n \pi(2k - \nu)$.
Thus, in this resonant case, the rotation is always a multiple of a half-circle rotation, which reduces the usable phase space (and therefore the reservoir computing performance) and leads to the resonance lines of bad performance found in Fig. \ref{fig:res_plot}.

\subsection{Hardware implementation}\label{Sec:3.4}

There have been various hardware implementations of photonic delay-based reservoir computing showing good performance on a range of benchmarking tasks. We refer the reader to \cite{SAN17a} for a general review of photonic reservoir computing, \cite{BRU18a} for a review specifically about delay-based photonic reservoir computing, and \cite{TAN19c} for a general review on the hardware implementation of reservoir computing, including time-delayed systems. Here we shall mention only selected papers.

Opto-electronic and all-optical implementations of delay-based reservoir computing have demonstrated good performance in benchmarking tasks such as spoken digit recognition and non-linear channel equalisation \cite{PAQ12,BRU13a,DEJ14}, at data processing rates up to 1\,Gbyte/s \cite{BRU13a}. However, in these experimental setups, results have only been reported for fixed feedback delay-times that are either equal to the clock-cycle \cite{BRU13a} or chosen according to the so-called desynchronised setup with $\tau=T+\theta$ \cite{PAQ12,DEJ14}. The results of the theoretical studies discussed in Section \ref{Sec:res} indicate that further improvement in the performance could be achieved by freely tuning the feedback delay-time. 

In \cite{TAK18} a reservoir consisting of a semiconductor laser with feedback from a short external cavity is studied. In this work, the authors aim to increase the number of virtual nodes under the constraint of viewing the reservoir as a network, meaning that they restrict the clock-cycle to $T=n\tau$ for $n\in\mathbb{N}$. This restriction, which is unnecessary for the time-multiplexed reservoir computing concept, is detrimental to the memory capacity of the reservoir. There are two factors that lead to decreased memory capacity when $T=n\tau$ for $n\in\mathbb{N}$. Firstly, the resonance between $T$ and $\tau$ decreases the memory capacity, as discussed in Section \ref{Sec:res} and shown in \cite{STE20}. Secondly, with each roundtrip in the feedback cavity the influence of past inputs decreases. Therefore, as $n$ is increased, the memory of previous clock-cycles decreases, as is demonstrated by the low memory capacities above the main resonance ($T=\tau$) in Fig.~\ref{fig:res_plot}a. The authors of \cite{TAK18} mitigate the low memory of their setup by including multiple past inputs in the input signal. A similar approach of using multiple delayed inputs was recently investigated theoretically in \cite{JAU21a}, showing promising results for task-specific performance enhancement  in hardware implemented reservoirs.

\section{Reservoir computing with multiple delays}\label{sec:4}

An extension of the delay-based reservoir computing setup reviewed in the previous section is to include additional delay-lines. So far, there have only been a few works on this topic which are mostly theoretical. However, one of the earliest works was an experimental study using an opto-electronic reservoir \cite{MAR12a}. In \cite{MAR12a} 15 time-delayed feedback lines were included. The delay-times were chosen as integer multiples of the virtual node separation, $\tau_n=n\theta$ for $n\in\mathbb{N}$, with the largest delay being equal to the clock-cycle, $\tau_{\textrm{max}}=T=N_v\theta$. The feedback weights were randomly selected and the equivalent network had sparse connectivity. The authors described their approach as a method of providing enhanced dynamical connectivity within a network view of the reservoir. Once again, the restrictions placed on the feedback delay-times in the setup used in \cite{MAR12a} do not allow for tasks specific delay-tuning of the memory capacity.

In \cite{NIE17} a theoretical investigation of the influence of a second time-delayed feedback term is investigated. The first delay is chosen resonant with the clock-cycle, $\tau_1=T$, and the second is given by $\tau_2=n\theta$ for $n\in\mathbb{N}$, where $n$ is varied to find task specific optimal values. At resonances between $\tau_1$ and $\tau_2$, the authors find poor performance for the NARMA10 task. This holds with the results of \cite{STE20} since $\tau_1=T$ in \cite{NIE17}. 

In \cite{HOU18} a semiconductor laser with dual optical feedback is simulated. For the first delay-line the desynchronised scheme, $\tau_1=T+\theta$, is chosen and the second delay is tuned. In this work, a strong dependence on $\tau_2$ is shown for the Santa Fe time series prediction task, as well as resonance effects between $\tau_1$ and $\tau_2$. Similarly, in \cite{CHE19c} an opto-electronic dual feedback scheme is simulated, also with $\tau_1=T+\theta$ and tuned $\tau_2$. Here a strong $\tau_2$ dependence and delay resonance effects are also demonstrated.

In each of the above mentioned studies \cite{MAR12a,NIE17,HOU18,CHE19c}, at least one of the delay-times is restricted to $\tau=T$ or $\tau=T+\theta$. Furthermore, the ranges over which the second delay-time is tuned do not exceed $4T$ in these studies. However, task specific memory requirements could demand much larger delays, as will be demonstrated for the NARMA10 task in Section \ref{Sec:new} and as is indicated by the task-specific dependence of the performance on delayed inputs recently shown in \cite{JAU21a}.

\section{Reservoir computing with multiple delay-coupled nodes}\label{sec:5}

So far we have considered delay-based reservoirs consisting of one non-linear node with feedback, with the multiplexing done in time. Hybrid spatial- and time-multiplexing approaches can also be realised, for example by delay-coupling multiple non-linear nodes. The advantage of such an approach is that the output dimension can be increased without the need for increasing the clock-cycle ($\theta$ can not be made arbitrarily small due to the finite time needed for a dynamical system to respond to the input). There are a number of theoretical studies on this topic \cite{ORT17a,ROE18a,GUO19,HOU19,SUG20,LIA21}, however in none of these studies is the influence of the coupling delay-times considered as a tunable parameter, rather the focus is on increasing the output dimension or investigating the influence of system parameters other than the delay-times.  

In \cite{ORT17a} a comparative study is carried out on the performance of two delay-coupled non-linear nodes and two uncoupled non-linear nodes with self-feedback where the outputs are collected into the same state vector. Various parameters of their chosen nonlinearity, as well as input and feedback scaling parameters, are scanned to find areas of optimal performance for various benchmarking tasks. However, the delay-times are kept fixed and equal to the clock-cycle. The authors find that it is both parameter and task dependent as to whether the performance is better in the coupled or the uncoupled system. 

 Variants of two semiconductor lasers delay-coupled in a ring configuration are considered in \cite{HOU19,GUO19,LIA21}, with either resonant delay-times ($\tau=T$) or the desynchronised case ($\tau=T+\theta$). In \cite{HOU19} the influence of slightly varying one of the delay-times is considered, however, only over a small range with the aim of showing that the performance does not degrade if it is not possible to make the two delays exactly equal in an experimental setting. 
 
 In \cite{SUG20} the authors study the performance of multiple uncoupled semiconductor lasers with self-feedback, where, as in \cite{ORT17a}, the outputs are collected into one state vector. The purpose of this work is to demonstrate a method of increasing the output dimension. All delay-times are chosen equal and resonant to the clock-cycle in this study.
 
 Finally, in \cite{ROE18a} the effect of mixing spatial- and time-multiplexing was investigated. The authors set the total output dimension to a fixed value and varied the ratio of real and virtual nodes. They considered various delayed ring-coupling configurations for the real nodes. For all configurations, they found optimal performance for the NARMA10 task when the number of virtual nodes is larger than the number of real nodes. In this study the number of delay-lines and the length of the delay-lines varied with the proportion of virtual to real nodes, however, this was done to keep the ratio of the number of virtual nodes and the delay-times constant rather than to investigate the influence of varying delay-times.
 
 As of yet, the potential performance enhancement that could be achieved by introducing multiple time scales via the delay-times, in multi-node systems, has not been thoroughly investigated.
 
\section{New insights into memory capacity tuning with multiple delays}\label{Sec:new}

In this section, we present new sights into how a second time-delay influences the memory capacity and how tuning this delay can lead to improved performance for the NARMA10 task. We do this using a system of two coupled Stuart-Landau oscillators, however, some of these results are also applicable for a reservoir consisting of a single non-linear node with multiple feedback terms.  

\subsection{Coupled Stuart-Landau oscillators}

To investigate the influence of multiple delays and coupled nodes, we use a system of two delay-coupled Stuart-Landau oscillators. The Stuart-Landau oscillator is a generic system describing the behaviour near a Hopf bifurcation and is therefore suitable for investigating the influence of delayed coupling and delayed feedback in a general context. The corresponding equations of motion are

\begin{align}\label{eq:stuart_landau_ring_und_self}
  \dot{z}_{j}=\left(\lambda_{j} + \eta_{j} g_{j}(t)u(t) + i\omega_{j} + (\gamma_{j} + i\alpha_{j})|z_{j}|^2\right) z_{j} ~~ + ~~\kappa_{j}~ e^{i\phi_{j}} ~z_{j+1}\left(t-\tau_{j}\right)
    ~+ ~~~~~~~~~~~~~~~~\nonumber\\ \kappa^{self}_{j}  e^{i\phi^{self}_{j}}z_{j}\left(t-\tau^{self}_{j}\right) +~ D\xi(t).
\end{align}

\begin{figure}[th]
\centering\includegraphics{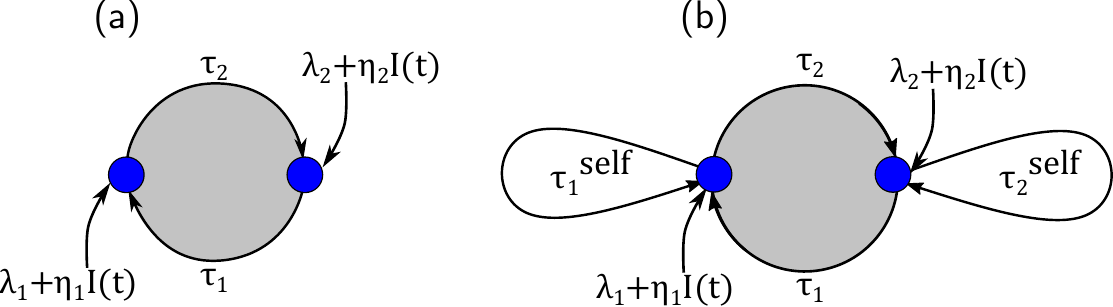}
\caption{Scheme of two oscillators (blue circles) delay-coupled in a ring-topology without (a) and with (b) self-feedback. The coupling is unidirectional, with coupling strength $\kappa_j$ and delay $\tau_j$ ($j \in \left[1,2\right]$). Both nodes are externally driven with constant pump rate $\lambda_j$ and time dependent input $I_j(t)= g_j(t)u(t)$ scaled with $\eta_j$.}
    \label{fig:ring_scheme}
\end{figure}

\noindent Here, the index $j$ labeles the oscillator $j \in \left[1,2\right]$, $\lambda_j$ denotes the pump rate, $\omega_j$ the frequency, $\gamma_j$ the nonlinearity parameter, $\alpha_j$ the re-scaled sheer, $\kappa_j$ the feedback strength, $\phi_j$ the feedback phase and $\tau_j$ the delay-time. Optional self-feedback is included with the self-feedback strength, phase  and delay-time given by $\kappa^{self}_j$, $\phi^{self}_j$ and $\tau^{self}_j$, respectively. $\xi(t)$ models Gaussian white noise with amplitude $D=10^{-8}$. 
The input is denoted as $\eta_j g_j(t)u(t)$, where $u(t)$ is the original (unmasked) input sequence, $g_j(t)$ denotes the $j^\textrm{th}$ mask function and $\eta_j$ the input strength. The mask functions were chosen to be piecewise constant random binary. It is important to note that each oscillator $j \in \left[1,2\right]$ has a distinct mask. Figure \ref{fig:ring_scheme}a illustrates the delayed, ring-coupling topology, and Fig.~\ref{fig:ring_scheme}b shows the ring-coupling with self-feedback. 

\subsection{Simulation details}\label{sec:simulation_details}
Simulations were done in C++ using standard libraries and the linear algebra library "Armadillo" \cite{SAN16}.
The delayed Stuart Landau equations Eq.~\eqref{eq:stuart_landau_ring_und_self} were integrated using a Runge-Kutta 4th order algorithm with $\Delta t=0.01$. The matrix inverse was calculated via the Moore-Penrose-Pseudoinverse from the C++ linear algebra library "Armadillo".
For all simulations, the reservoir was initialised by first simulating without input for an integration time of 1000$\tau_1$ and then simulating with input for a buffer time of 1000 input clock-cycles.
In the training stage, 5000 input values were used for the NARMA10 task and 40000 for the memory capacity. In both cases, the testing procedure involved 5000 inputs.
Before feeding the input into the reservoir, it was normalised via a linear transformation, such that $u(t)\in [-1,1]$. This increases comparability between the memory capacity calculations and NARMA10, and ensures that inputs do not become too large. To avoid over-fitting, for the NARMA10 task regularisation by noise was included. This means that the Tikhonov regularisation parameter introduced in Eq.~\eqref{Eq:wout} was set to $\lambda=0$ and instead all entries of the state matrix were disturbed by additive Gaussian white noise with standard deviation $D=10^{-8}$. Regularisation by noise has been shown to be equivalent to Tikhonov regularisation \cite{BIS95b}. For the masks, random binary values in $\{0,1\}$ were chosen. The default parameters used for simulation are given in table \ref{tab:default_parameters}.

\begin{table}[htpb]
    \centering
    \caption{Default dimensionless parameter values for the numerical simulations of the reservoir computing performance of two oscillators (index $j$) described by Eq.~\eqref{eq:stuart_landau_ring_und_self}.}
    \begin{tabular}{l l l l l l} 
    \hline
         Parameter & Description & value &Parameter& Description & value  \\ 
         \hline
          \vspace{-1mm} $\tau_j$ & Delay &  425 & $T$ & Clock cycle & 200\\ \vspace{-1mm}
         $\lambda_j$ & Pump rate & 0.01 & $\kappa_j$ & Feedback strength & 0.18\\ \vspace{-1mm}
         $\phi_j$ & Feedback phase & 0 &  $\eta_j$ & Input strength &  0.06\\ \vspace{-1mm}
          $\alpha_j$ & Sheer parameter &  0 & $\omega_j$ & Frequency & 0\\ \vspace{-1mm}
        $\gamma_j$ & Nonlinearity &  -0.1 & $N_v$ & Virtual nodes per oscillator & 100\\ 
       \hline
    \end{tabular}
    \label{tab:default_parameters}
\end{table}

\subsection{Coupling-delay and clock-cycle resonances in a ring-coupled system}\label{sec:resonances}

In this section, we will investigate the influence of a second feedback delay-time in a system of two ring-coupled nodes by allowing the delay-times $\tau_1$ and $\tau_2$ to be different. This setup corresponds to Fig.~\ref{fig:ring_scheme}a, i.e. Eq.~\eqref{eq:stuart_landau_ring_und_self} with $\kappa^{self}_{1,2}=0$.

\begin{figure}[t]
\centering\includegraphics[width=0.65\textwidth]{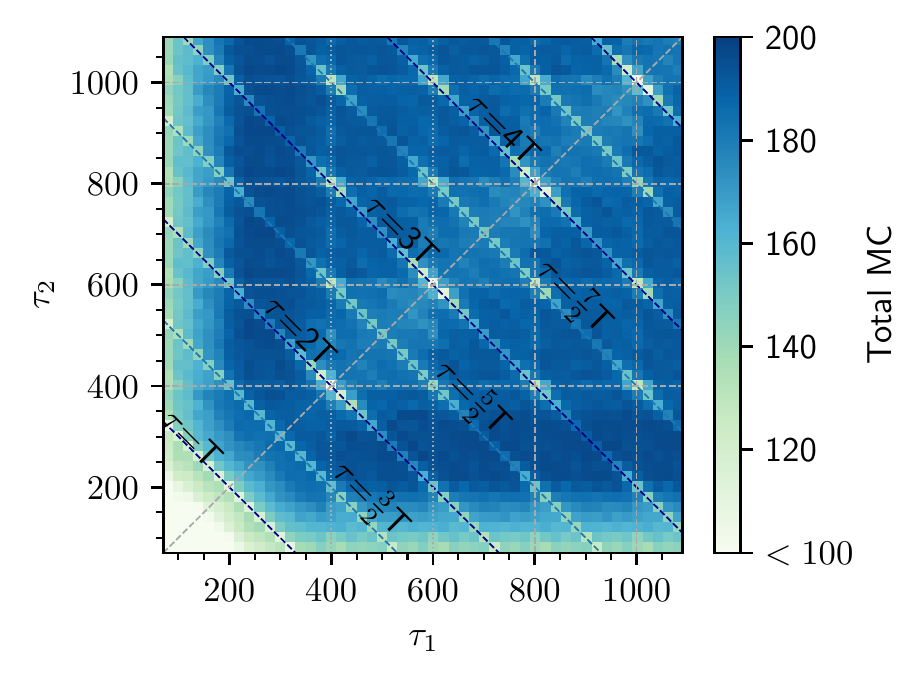}
\caption{Total memory capacity of two mutually-coupled oscillators (see Fig.\ref{fig:ring_scheme}a and Eq. \eqref{eq:stuart_landau_ring_und_self} with $\kappa_j^{self}=0$), color coded as a function of the ring-delays $\tau_1$ and $\tau_2$. Dashed lines indicate $1/2(\tau_1+\tau_2)=n/m T$ (mean delay resonance-lines) with $n,m \in \mathbb{N}$, $\tau_{1,2}=n/m T$ and $\tau_1=\tau_2$, respectively. Resonances are especially pronounced at the intersections where both delays are in resonance.}
    \label{fig:resonanz_tau_mismatch}
\end{figure}

As discussed in Section~\ref{Sec:res}, previous works show that resonances between the delay-time $\tau$ and the clock cycle $T$ decrease the memory capacity in systems with one delay \cite{STE20,KOE20a}. These resonances occur at $m\tau=nT$ for $n,m\in \mathbb{N}$ (see Fig.~\ref{fig:res_plot}a). Introducing a second delay via the ring-coupling scheme gives the same resonance structure when $\tau_1=\tau_2$. However, if we introduce a mismatch between the delays, i.e. $\tau_1\neq \tau_2$, resonances are found at rational multiples between the mean delay, $\tau_\textrm{mean}=\left(\tau_1+\tau_2\right)/2$, and the clock cycle: $m\tau_\textrm{mean}=nT$ for $n,m\in \mathbb{N}$. This resonance structure is shown in Fig.~\ref{fig:resonanz_tau_mismatch}. The resonances are especially pronounced at the points, where both delays are in resonance with the clock cycle (intersection points of horizontal and vertical lines in Fig.~\ref{fig:resonanz_tau_mismatch}), with the largest losses in the memory capacity being found at the resonances points where both delays are equal, i.e. $\tau_1=\tau_2=nT$ for $n\in\mathbb{N}$.

In order to understand why the resonances occur at the mean delay, it is useful to show how the delay-mismatched system can be rewritten in terms of a system with two identical delays. 
A system of $N$ oscillators in a delayed ring-topology can generically be described by
\begin{align}
    \dot{z}_j&=f(z_j(t),I_j(t))+ g(z_{j+1}(t-\tau_j)),\label{eq:ring_input}
\end{align}
where $I_j(t)=\eta_j g_j(t)u(t)$ is the masked injection function, $u(t)$ is the original (unmasked) input, $f(\cdot)$ is the nonlinearity and $g(\cdot)$ is the coupling term. For the Stuart-Landau oscillator system one would have $f(z_j,I_j)=\left(\lambda_j + I_j (t) + i\omega_j + (\gamma_j + i\alpha_j)|z_j|^2\right) z_j$ and $g(z_{j+1}(t-\tau_j))=k_j e^{i\phi_j}g(z_{j+1}(t-\tau_j))$.
Note, that the ring-topology implies $z_{N+1} \equiv z_1$.
Equation~\eqref{eq:ring_input} can be transformed into a system of $N$ delay-coupled oscillators with identical delays via the following transformation to a new coordinate $v_j$ \cite{PER10c}:
\begin{align}
    v_j(t)=z_j(t-t'_j),~~\textrm{with}~~t'_{j}=(N-j)\cdot\tau_\textrm{mean}-\sum^{N-j}_{i=1}\tau_{N-i},
\end{align}
where $\tau_\textrm{mean}=\sum_i \tau_i/N$ denotes the mean delay. This leads to a system of delay-differential equations for the transformed coordinate $v_j(t)$ with identical time-delays:
\begin{align}\label{eq:transformed_equivalent_sys}
    \dot{v}_j=f(v_j,I(t-t'_j)) + g(v_{j+1}(t-\tau_\textrm{mean})).
\end{align}
This transformation shows that the ring-coupled system with mismatched delays is equivalent to a ring with identical delays, equal to the mean delay, however, with delayed input $I(t-t'_j)$. In the equivalent system, we expect resonances with the clock cycle $T$ for $\tau_\textrm{mean}=\frac{\sum_i\tau_i}{N}=\frac{m}{n} T;~m,n \in N$. Consequently, these are the resonances found in the original system (see Fig.~\ref{fig:resonanz_tau_mismatch}).

For the equivalent system, Eq.~\eqref{eq:transformed_equivalent_sys}, the input term is shifted from $I_j(t)$ to $I_j(t-t'_j)$. This implies a general correspondence between ring-coupled systems with multiple delays and systems with delayed inputs. In \cite{JAU21a} it was recently shown that an appropriately chosen additional delayed input can improve the reservoir computing performance of a delay-based reservoir for various benchmark tasks \cite{JAU21a}. In order to show how a similar effect can be achieved in the ring-coupled system, it is necessary to look at the dependence on the delay-times of the individual contributions to the total memory capacity. 

\begin{figure}[t]
\centering\includegraphics[width=\textwidth]{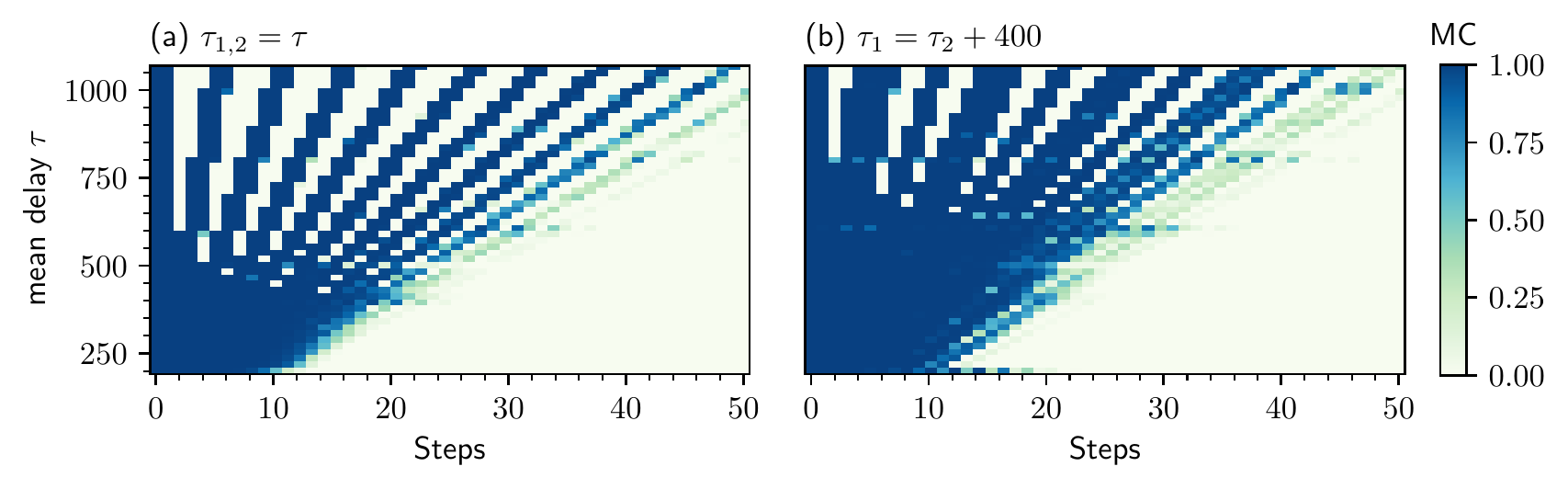}
\caption{Impact of delay-mismatch: Linear memory capacity $C_{u_{-steps}}$,  color coded as a function of mean delay and steps into the past, averaged over 10 masks. (a) Identical delays $\tau_1=\tau_2$, (b) Mismatched delays $\tau_1=\tau_2+400$,
}
\label{fig:bigplot}
\end{figure}

Figure~\ref{fig:bigplot} shows the linear memory capacity as a function of the mean delay and the number of steps into the past. In agreement with previous work, increasing the delay-time of a reservoir while keeping the clock cycle constant increases long-time memory capabilities \cite{KOE20a}. However, as the delay-times become much larger than the clock-cycle and the characteristic time scale of the non-linearity, gaps in the memory capacity begin to form. 
More precisely, the gap for input $u_{-n}$ occurs at $\tau\geq (n+1)T$, since if the $t_0$ is the time the current input begins to be fed into the system, then $u_{-n}$ was fed into the system in the interval $[t_0-nT,t_0-(n+1)T[$. Note, that due to the non-zero relaxation time of the dynamical nonlinearity, the gaps in the memory capacity may be shifted to slightly higher delay-times. 
These gaps can be reduced by mismatching the delay-times, as can be seen by comparing Fig.~\ref{fig:bigplot}a, where $\tau_1=\tau_2$, with Fig.~\ref{fig:bigplot}b, where $\tau_1=\tau_2+400$. For increasing mean delays, more long-time memory contributions are present, with smaller gaps in the mismatched case. This phenomenon is valid for the linear and non-linear capacities (corresponding scans for the quadratic terms are shown in the supplemental material). In \cite{GOL20} a similar reduction of gaps in the memory capacity was achieved by coupling systems with multiple delays. In contrast to our contribution, the authors used a so-called deep reservoir computing scheme. With our setup, we are able to achieve similar results in the less complex ring-coupling scheme. 

The filling of the gaps in the memory capacity for mismatched delays can be understood qualitatively. We first note that, even if $\tau_\textrm{mean}=0$ the system without delay has some capability to memorise inputs short distances into the past, due to the non-zero relaxation time of the dynamical nonlinearity. For the sake of simplicity, this contribution is neglected in the following argumentation.
When a delay is introduced, each delay line $\tau_j$ "stores" information about previous inputs and possibly nonlinear transformations of them. For instance, if we consider a ring of two oscillators with $\tau_1=\tau_2=3T$, then the ring will store information about inputs $u(t-3T),u(t-6T),u(t-9T)$ and so on. Since each input is fed into the system for the duration $T$ this corresponds to $u_{-3},u_{-6},u_{-9},...$, clearly leaving some gaps of previous inputs, that cannot be remembered. On the other hand, if we choose a system with equal mean delay, but a mismatch in the delay, e.g. $\tau_1=2T\tau_2=4T$, we obtain information about $u_{-2},u_{-4},u_{-6},u_{-8}$ and so forth, filling some of the gaps in the equal delay case. 

Aside from being able to fill gaps in the memory capacity by mismatching the delays, Fig.~\ref{fig:bigplot} shows that the delays can be chosen such that the system fulfils particular memory requirements that could be beneficial for specific tasks. However, in the ring-topology, varying one of the delay-times always influences the mean delay, meaning that the effect of large differences in the two delays can not be fully utilised. Therefore, in Section~\ref{sec:delay_time_manipulation} we include an additional self-feedback term that allows for more tunability of the memory of the system.

\subsection{Manipulating the range of accessible past inputs via additional feedback lines}\label{sec:delay_time_manipulation}

In this section, we show how an additional self-feedback term can be used to tune the memory capacity of our system of two ring-coupled nodes and how this can be used to improve the performance of the NARMA10 task. This setup corresponds to Fig.~\ref{fig:ring_scheme}b, i.e. Eq.~\eqref{eq:stuart_landau_ring_und_self} with $\kappa^{self}_{1,2}\neq 0$. For simplicity, we choose parameters identical to the default ring-topology parameters for all nodes and set the self-feedback coupling parameters to the values of the ring-coupling. The self-delays are chosen equal,  $\tau^{self}=\tau_1^{self}=\tau_2^{self}$, and are treated as a free parameter. By choosing $\tau_1^{self}=\tau_2^{self}$ and $\tau_1=\tau_2$, there are effectively two independent time-scales introduced by the delays. This setup is similar to a single nonlinear node with two delayed-feedback terms, therefore all delay-dependent effects shown below are also applicable to such a system.

\begin{figure}[t]
\centering\includegraphics[width=\textwidth]{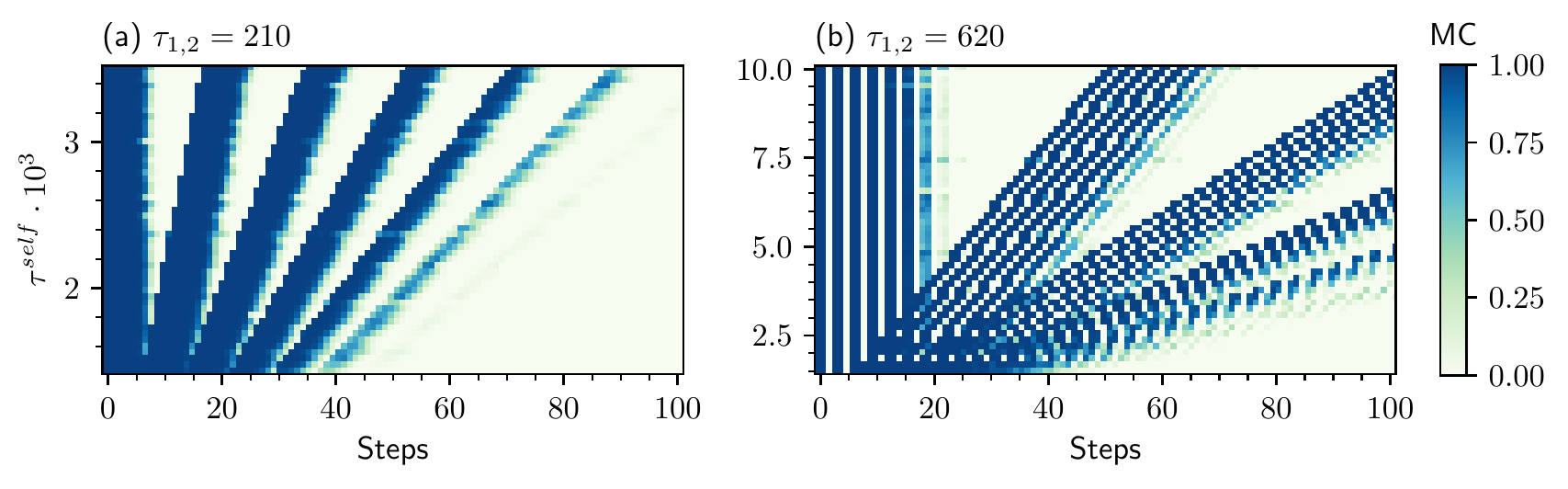}
\caption{Impact of self-delay: Linear memory capacity $C_{u_{-steps}}$  color coded 
as a function of the self-delay $\tau^{self}$ and the steps into the past, averaged over 10 mask realisations for (a) $\tau_{1}=\tau_{2}=210$, and  (b) $\tau_{1}=\tau_{2}=620$.
}
\label{fig:bigplot2}
\end{figure}

In this setup, the range of past inputs that the system can recall can freely be altered.
Figure~\ref{fig:bigplot2}a shows the linear memory capacities, as a function of the number of steps into the past, over the self feedback time $\tau^{self}$, while fixing the delay-time of the ring-coupling. We obtain different regions of past inputs, the system can remember. First, there is a region of inputs that corresponds to the memory capabilities of the ring-coupling. This region does not change with increasing self-delay and therefore appears as a vertical stripe. Second, there are additional regions of past inputs the system can remember that correspond to multiples of the self-delay time. Their position is completely determined by the value of the self-delay, $\tau^{self}$ and can therefore be tuned as desired via the value of the self-delay. These regions appear as oblique beams in the Figure~\ref{fig:bigplot2}. In the supplemental material, we show, that the method works for higher order terms, i.e. the quadratic capacities as well. 

Between the regions of high memory capacity, there are gaps at steps that cannot be recalled.
The width of the regions that can be remembered, can be tuned via the value of the ring-delay. Higher ring-delays yield larger width of the memorisable areas, whereas smaller values yield smaller width. Nevertheless, the manipulation of the width is limited, because high ring-delays also yield gaps in the memory capacity (see Fig.\ref{fig:bigplot}a). Such a case is depicted in Fig.~\ref{fig:bigplot2}b, where a ring-delay above the threshold, where gaps in the memory capacity begin to occur, was chosen. Note, that for fundamental reasons, a reservoir computer cannot remember all previous inputs (fading memory property). It is therefore only possible to tune which ones contribute to the total memory capacity.

\begin{figure}[h]
\centering\includegraphics[width=\textwidth]{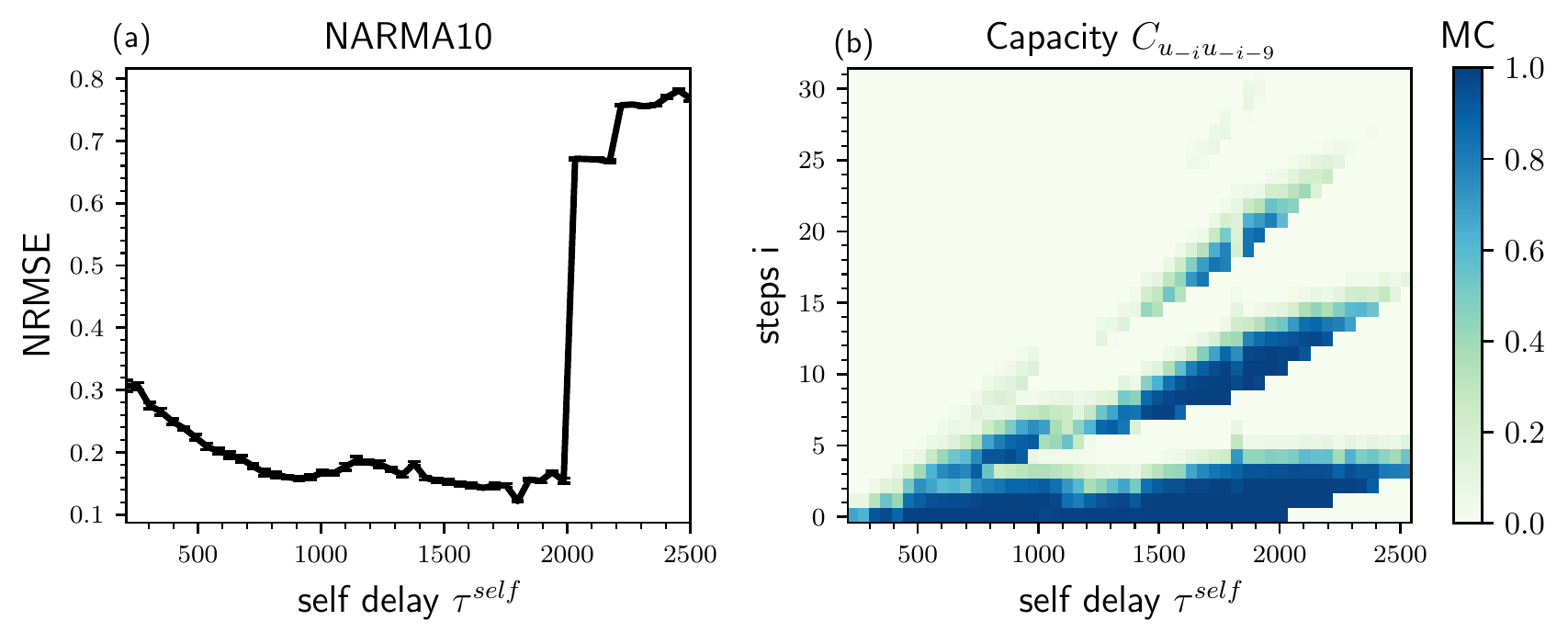}
\caption{(a) NARMA10 computing error (NRMSE) as a function of the self-delay time $\tau^{self}$ in the all-to-all coupled topology displayed in Fig.\ref{fig:ring_scheme}b, averaged over 10 mask realisations (error bars indicate standard deviation). (b) Memory capacity corresponding to input terms $u_{-i}u_{-i-9}$ as they occur in the NARMA10 series Eq. \eqref{eq:NARMA10} as a function of the self delay-time and the recall step. }
\label{fig:NARMA_NRMSE_STEPS}
\end{figure}

So far we only considered memory capacities and how they can be influenced by tuning the delay-times of the system. We will now investigate the connection between memory capacities and a specific task, and show that the delay-times can have a significant influence on the task specific performance. As an exemplary task we use NARMA10. In Fig.~\ref{fig:NARMA_NRMSE_STEPS}a the NARMA10 error (NRMSE) is shown as a function of the self-delay. As can be seen, if the delays are chosen equal, $\tau^{self}=\tau_{1}=\tau_2$, the NRMSE starts with moderate values of around 0.3. Increasing the self-delay, the NRMSE decreases to slightly above 0.1. This is to our knowledge unprecedented in the reservoir computing literature for this number of virtual nodes (200) and demonstrates that the delay-times are an important tuning parameter for improving task specific performance. 

Further increasing the self-delay above $\tau^{self}=2000$, which is 10 times the clock cycle, leads to a sudden performance drop-off. With the previously obtained knowledge of the influence of delays on the memory capacity, we can understand this behaviour. In order to explain the effect, consider the capacities that correspond to the input terms $u_{-i-9},u_{-i}$ that appear in the NARMA10 time series Eq.~\eqref{eq:NARMA10}. 
In Fig.~\ref{fig:NARMA_NRMSE_STEPS}b we plot the capacities $C_{u_{-i-9},u_{-i}}$ corresponding to the mentioned terms over the value of the self-delay time.
As one can see, the capacity corresponding to these terms first increases with increasing $\tau^{self}$. Then for too high self-delays, the system cannot remember the first of the relevant terms and therefore the system's performance drops dramatically. This is due to the fact that, for these high self-delays, the memory capacity gets additional gaps at timesteps, that are needed for the NARAM10 task.

\section{Conclusion}

In this article, we have reviewed delay-based reservoir computing with a focus on the influence of feedback and coupling delay-times. We aim to bring to the attention of the wider reservoir computing community that the delay-times have the potential to play a far more important role in improving the performance of delay-based reservoir computers than they have up until now. Furthermore, we highlight that, although it was important for the conception for delay-based reservoir computing, there is no need to view delay-based reservoirs as networks. Viewing delay-based reservoirs as networks has, in fact, led to a common misconception about the relationship between the clock-cycle and the delay-times of the reservoir. By introducing the concept of time-multiplexed reservoir computing independent of time-delayed systems, we demonstrate that there is no predetermined relationship between the clock-cycle and the reservoir delay-times. 

Our results underline, in agreement with recent studies, that choosing the delay-time resonant with the clock-cycle is generally detrimental to the computing capabilities of the reservoir. Furthermore, also the commonly used desynchronised case, where delay and clock cycle only differ by one virtual node, does not lead to task specific optimal performance. 

Additionally, we have demonstrated the influence of a second delay-time by considering a system of two delay-coupled oscillators. We have shown how varying these delay-times influence the linear memory capacity and how the addition of a second delay can fill in gaps in the linear memory capacity that appear in single delay systems. We showed that in a ring-coupled configuration, resonances between the delays and the clock-cycle depend on the average delay-time. Further, the influence of the delays on the memory capacity can be tuned more freely when the added delay-time is introduced as a self-feedback term. In this case, the delay-times can be independently tuned to achieve the desired linear memory capacity, including both long- and short-term memory. For a task such as NARMA10, which has very specific memory requirements, correctly tuning the delays leads to a significant improvement in the performance.

\begin{backmatter}

\bmsection{Acknowledgments}
We acknowledge support be the Deutsche Forschungsgemeinschaft(DFG) in the framework of the SFB910.

\bmsection{Disclosures}
\noindent The authors declare no conflicts of interest.

\bmsection{Data Availability Statement}
\bmsection{Data availability} Data underlying the results presented in this paper are not publicly available at this time but may be obtained from the authors upon reasonable request.

\bmsection{Supplemental document}
See Supplement 1 for supporting content. 

\end{backmatter}

\bibliography{ref}

\end{document}